\documentstyle[aps]{revtex}
\begin{document}
\title{Excitonic effects in solids described by time-dependent density
functional theory }

\author{ Lucia Reining$^{1}$, Valerio Olevano$^{1}$, Angel Rubio$^{1,2}$ and
Giovanni Onida$^{3}$}

\address{1) Laboratoire des Solides Irradi\'es, CNRS-CEA,
          \'Ecole Polytechnique,
          F--91128 Palaiseau, France \\
      2) Departamento de F\'\i sica de Materiales,
         Facultad de Ciencias Qu\'\i micas,
     Universidad del Pais Vasco, \\
     and  Donostia International Physics Center.
     E--20018 San Sebasti\'an, Basque Country, Spain \\
      3) Istituto Nazionale per la Fisica della Materia, \\
          Dipartimento di Fisica dell'Universit\`a di Roma ``Tor
Vergata,''
          I--00133 Roma, Italy}


\date{\today}

\maketitle
\begin{abstract}
Starting from the many-body Bethe-Salpeter equation we
derive an exchange-correlation kernel $f_{xc}$ that reproduces
excitonic effects in bulk materials within time-dependent 
density functional theory.
The resulting $f_{xc}$ accounts for both
self-energy corrections and the electron-hole interaction.
It is {\em static}, {\em non-local} and 
has a long-range Coulomb tail. Taking the example of
bulk silicon, we show that the $- \alpha / q^2$ divergency is
crucial and can, in the case of continuum excitons, even be sufficient for
reproducing the excitonic effects and yielding excellent agreement
between the calculated and the experimental absorption spectrum.
\end{abstract}

\newpage

\twocolumn

The calculation of electronic excitations has remained a
major challenge in the field of solid state theory. In fact, whereas
ground state properties can be
computed today with good precision within
Density Functional Theory (DFT) \cite{HKS},
electronic excitations are accessible only through additional
corrections. Within Many-Body Perturbation Theory, Hedin's GW
corrections \cite{H} are used to get electron addition and
removal energies, and the Bethe-Salpeter Equation (BSE) for neutral 
excitations as those measured
for example in absorption or electron energy loss spectroscopy
(EELS).
In fact, since several years {\it ab initio} BSE calculations
yield generally good
agreement between the calculated and the experimental 
absorption spectra for
both finite \cite{ORGDA} and infinite systems \cite{ARDO,BBS,RL}.
However these calculations are
necessarily cumbersome, which has up to now prevented their
application to very complex systems.

In the last years, an alternative approach has been developed which
allows in principle to correctly describe exchange-correlation effects
in the neutral excited state, namely Time-Dependent
Density Functional Theory (TDDFT) \cite{RGK,GDP}.
As in the case of static
DFT the main obstacle resides in finding a good approximation to the
unknown exchange-correlation (xc) contribution. For the ground state
properties of the majority of finite and infinite systems, the
local density approximation (LDA) has
turned out to yield surprisingly good results. 
In the case of the absorption and EELS spectra of finite 
systems, and of EELS spectra of solids,
TDDFT has yielded good results using the 
adiabatic LDA approximation (TDLDA)
for the xc-kernel $f_{xc}$\cite{GDP}. 
However, this is not true for the {\it absorption} spectra
of {\it solids}\cite{tdlda}.
In fact, it has turned out that the results for the latter obtained
within TDLDA are extremely close to those obtained in a simple
Random Phase Approximation (RPA) calculation, where the
xc-kernel is completely neglected, and only local
field effects (in other words, the contribution coming from the
variation of the Hartree potential) are taken into account. It should
be pointed out that this similarity between RPA and TDLDA spectra
to some extent also holds in the case of clusters and EELS spectra of
solids, but the results are nevertheless satisfactory because 
in those cases already RPA spectra, including local field effects,
are in good agreement with experiment.

On the other hand, it would be extremely desirable to obtain good
absorption spectra of solids within TDDFT,
since the equations to be solved are two-point ones, in contrast to
the four-point BSE.  The obstacle to be removed to this aim is
hence the fact that a good approximation to the
xc-kernel must be found. In this context, extensive
discussions can be found in literature about the need to include
long-range nonlocal terms and dynamical (memory) effects in the kernel
\cite{GDP,TokatlyPankratov}.

In this work, we show how a TDDFT equation for the
macroscopic dielectric function can be derived from the
Bethe-Salpeter equation. The derivation is exact in the sense that the
resulting equation does yield the same spectra as
the standard Bethe-Salpeter equation (with its own various approximations
\cite{ARDO,BBS,RL}).
We demonstrate that the
resulting xc-kernel has a $1/q^2$ contribution, the strength of which
is inversely proportional to the screening in the system. Finally, we
calculate the absorption and the refraction index spectra of bulk silicon
within TDDFT, using
only this {\it static} long range tail as xc-kernel, and obtain excellent
agreement with experiment.

To start with it is useful to put the
TDDFT equation and the Bethe-Salpeter equation for $\varepsilon^{-1}$
on the same footing. In fact, both equations can be written schematically
in the same Dyson-like form,
\begin{eqnarray}
   S = S^{(0)} + S^{(0)} K S. \label{SEQ}
\end{eqnarray}
Here, S can be either the two-point polarisability $\chi$, from which
we can obtain the inverse dielectric function
$\varepsilon^{-1} = 1 + v \chi$,
or $S=L$, the two-particle correlation function
which yields $\chi$ (and then $\varepsilon^{-1}$) by
contracting two of its four indices
\begin{eqnarray}
   \chi(x_1,x_2) = L(x_1,x_{1}^{+};x_2,x_{2}^{+}).
\end{eqnarray}
Here $x$ stands for space, spin and time coordinates.
Of course this 
holds only for what concerns the form, but not the specific details.
First, quantities in the TDDFT equation are two-point ones
whereas in the BSE they are four-point ones.
Second, in TDDFT $S^{(0)}$ is the independent-particle response function
$\chi^{(0)}$
constructed with the Kohn Sham (KS) orbitals and eigenvalues, whereas
in the BSE formalism $S^{(0)}$ stands for the independent
quasiparticle response $L^{(0)}$, i.e., it is constructed using
quasiparticle
eigenvalues and eigenfunctions (e.g., obtained from a GW calculation).
Finally, in the TDDFT case, the kernel $K$ is defined as
$K=v+F^{\rm TDDFT}$, where
$v$ is the Coulomb potential and $F^{\rm TDDFT}$ stands for 
the $f_{xc}$ kernel.
In the case of the BSE, $K=v+F^{\rm BSE}$,
where $v$ and $F^{\rm BSE}$ are the four-point functions
$v(x_{1},x_{2},x_{3},x_{4}) = \delta(x_1,x_2) \delta(x_3,x_4) v(x_1,x_3)$
and
$F^{\rm BSE}(x_{1},x_{2},x_{3},x_{4}) =
-\delta(x_2,x_4) \delta(x_1,x_3) W(x_1,x_2)$, where $W$ is the 
screened interaction.
In the case of TDDFT,
one can also understand Eq. (\ref{SEQ}) as a four-point equation, but
with
$F^{\rm TDDFT}(x_{1},x_{2},x_{3},x_{4}) =
\delta(x_{1},x_{2})\delta(x_{3},x_{4})
f_{xc}(x_{1},x_{3})$ which implies that one can immediately
contract the indices by pairs and reduce the equation to a two-point one.

If one performs
a basis transformation of the form $x_{l}\rightarrow \psi_{n}(x_{l})$,
where the $\psi_{n}(x_{l})$ are the
one-particle orbitals which diagonalise the
four-point $S^{(0)}$,
Eq. (\ref{SEQ}) 
can be transformed to an effective two-particle Hamiltonian equation, 
\begin{eqnarray}
{H}_{(n_{1}n_{2})(n_{3}n_{4})}^{2p} &\equiv&
\left(
\epsilon _{n_{2}}-\epsilon _{n_{1}}
\right)
\delta_{n_{1}n_{3}}\delta _{n_{2}n_{4}}
\nonumber \\
&+&
\left(
f_{n_{1}}-f_{n_{2}}
\right)
K_{(n_{1}n_{2})(n_{3}n_{4})}.
\label{Hexc def Bloch}
\end{eqnarray}
Here the indices $n_{i}$ refer to the fact that matrix elements involve
four eigenfunctions of the starting effective
one-particle Hamiltonian with eigenvalue $\epsilon_{n_{i}}$ and occupation
$f_{n_{i}}$. 
Defining the identity operator
$I=\delta_{m_{1}m_{3}}\delta_{m_{2}m_{4}}$,
$S$ is directly obtained from $H^{2p}$ as
\begin{equation}
  S_{(n_{1}n_{2})(n_{3}n_{4})}=\left[
  H^{{2p}}-I\ \omega \right]
  _{(n_{1}n_{2})(n_{3}n_{4})}^{-1}\left(
  f_{n_{4}}-f_{n_{3}}\right) .
\label{Chi Hexc}
\end{equation}
$H^{2p}$ can be diagonalized, and from
its eigenvalues $E_{\lambda}$ and
eigenstates $A_{\lambda}^{n_{1}n_{2}}$ the spectral representation of
$S$ can be constructed.

Since the
$\psi_{n}(x_{l})$ have to diagonalise $S^{(0)}$, they must be the KS orbitals
for the TDDFT equation and the QP eigenfunctions for the BSE. If these
functions are equal,  and if the  $A_{\lambda}$ and $E_{\lambda}$ are equal,
the BSE and TDDFT spectra would be the same. For a static $f_{xc}$ kernel, this
implies that the matrix elements of the Hamiltonians
$H^{2p}_{\rm TDDFT}$ and $H^{2p}_{\rm BSE}$ are equal.  In this scenario,
we can directly compare the BSE and TDDFT approaches.
First, in the
BSE the eigenvalues $\epsilon_{n}$ are quasiparticle energies (as
obtained for example from a GW calculation), whereas in TDDFT
they are the eigenvalues obtained from the
KS equation with the (in principle
exact) xc-potential. Second, for the BSE we have
\begin{eqnarray}
  F^{\rm BSE}_{(n_{1}n_{2})(n_{3}n_{4})} =
  - \int d{\bf r} d{\bf r'} \,
  \Phi(n_1,n_3;{\bf r}) W({\bf r'},{\bf r}) \Phi^{*}(n_2,n_4;{\bf r'}),
\nonumber
\end{eqnarray}
with the matrices $\Phi$ defined according to
\begin{eqnarray}
  \Phi(n_{1},n_{2};{\bf r}) := 
  \psi_{n_{1}}({\bf r})\psi^{*}_{n_{2}}({\bf r}),
\end{eqnarray}
whereas the exchange-correlation contribution to the TDDFT kernel
reads
\begin{eqnarray}
  F^{\rm TDDFT}_{(n_{1}n_{2})(n_{3}n_{4})} =
  \int d{\bf r} d{\bf r'} \, 
  \Phi(n_1,n_2;{\bf r}) f_{xc}({\bf r},{\bf r'}) \Phi^{*}(n_3,n_4;{\bf r'}).
\nonumber
\end{eqnarray}
(The Hartree contribution $v_{(n_1n_2)(n_3n_4)}$
is of course equal in both cases).
A comparison of the two cases does hence immediately tell us that the
BSE and TDDFT equations would yield the same spectrum
if the static $f_{xc}$ satisfies
\begin{eqnarray}
  &&
  (f_{n_{1}}-f_{n_{2}})
  \int d{\bf r} d{\bf r'} \,
  \Phi(n_1,n_2;{\bf r}) f_{xc}({\bf r},{\bf r'}) \Phi^{*}(n_3,n_4;{\bf r'})
  = 
  \nonumber \\ &&
  = {\cal F}_{(n_1 n_2)(n_3 n_4)},
\label{f versus W}
\end{eqnarray}
where\cite{footnote}
\begin{eqnarray}
  &&
  {\cal F}_{(n_1 n_2)(n_3 n_4)}
  =
  \left(
  \epsilon^{\rm QP} _{n_{2}}-\epsilon^{\rm QP} _{n_{1}} -
  \epsilon^{\rm DFT}_{n_{2}}+\epsilon^{\rm DFT} _{n_{1}}\right)
  \delta_{n_{1} n_{3}} \delta_{n_{2} n_{4}}\nonumber \\
  &&
  + (f_{n_{1}}-f_{n_{2}})
  F^{\rm BSE}_{(n_{1} n_{2}) (n_{3} n_{4})}.
\end{eqnarray}
It is clear that, if the transformation
$x_{l} \rightarrow \psi_{n}(x_{l})$
was complete in all four indices, Eq. (\ref{f versus W}) could never
be satisfied. 
The reason is that the two operators,
$F^{\rm TDDFT}$ and $\cal F$, {\em cannot be} equal 
because of the way the $\delta$ functions are put in real space.
On the other hand, if the two operators cannot be
made equal, then the spectrum can only be equal if at least one of the
two operators (in that case, $f_{xc}$) is energy dependent 
\cite{Wstatic}.
This in principle correct, general statement, can be made less
restrictive by realizing that in practice only 
a finite number of transitions
contributes to the optical spectrum. This means that we can use an incomplete
basis in transition space. 
In this reduced Hilbert space we can still find a static
operator that satisfies the required equality in transition space in a particular
energy range,
even though the real-space operators are {\it not} equal \cite{staticfxc}.
In order to discuss this possibility, we rewrite 
Eq. (\ref{f versus W}) as
\begin{eqnarray}
  &&
  (f_{n_{1}}-f_{n_{2}})
  \sum_{{\bf G},{\bf G'}} 
  \Phi(n_{1},n_{2};{\bf G}) 
  f_{xc}({\bf q},{\bf G},{\bf G'}) 
  \Phi^{*}(n_{3},n_{4};{\bf G'}) =
  \nonumber \\ &&
  = {\cal F}_{(n_1 n_2) (n_3 n_4)},
\label{fGG}
\end{eqnarray}
where ${\bf q} = {\bf k}_2-{\bf k}_1 = {\bf k}_4-{\bf k}_3 $.
Since for the particle-hole and hole-particle contributions of a
non-metal the factor $(f_{n_{1}}-f_{n_{2}})$ can never be zero, we
could in principle obtain the matrix $f_{xc}$ from
Eq. (\ref{fGG}) by inverting the matrices $\Phi$:
\begin{eqnarray}
  &&
  f_{xc}({\bf q},{\bf G},{\bf G'}) = 
  \sum_{n_{1} n_{2} n_{3} n_{4}}
  \frac{1}{(f_{n_{1}}-f_{n_{2}})}
  \nonumber \\ &&
  \Phi^{-1}(n_{1},n_{2};{\bf G})
  {\cal F}_{(n_1 n_2) (n_3 n_4)}
  (\Phi^{*})^{-1}(n_{3},n_{4};{\bf G'}).
\label{fxc}
\end{eqnarray}
%
%
As pointed out above, 
the sum over the indices $n_{i}$ is necessarily limited to some subspace
of important transitions, because otherwise the inverse
of $\Phi$ does not exist. In
certain cases, a subspace which at the same time allows the matrix to
be invertible and to reasonably reproduce the exciton spectrum
cannot be found. One example is a two-band model with $\epsilon^{\rm DFT}=
\epsilon^{\rm QP}$ consisting of plane
wave states $\psi_{b{\bf k}}({\bf r}) = e^{i{\bf G}_b {\bf r}}e^{i{\bf kr}}$, 
with ${\bf G}_b={\bf G}_v$ or ${\bf G}_{c}$.
In this case the matrices $\Phi$ are k-independent, which means
that all their rows are equal. In other words, the matrix
$F^{\rm TDDFT}$ is
k-independent, whereas the matrix $F^{\rm BSE}$ goes as
$1/(k-k')^{2}$, which is a clear contradiction.
On the other hand, if a subspace
where $\Phi$ is invertible can be found,
it is clear that the resulting kernel is
{\it frequency independent}, as none of the quantities implied, 
$F^{\rm BSE}$ or
$\Phi$, are frequency dependent (if the kernel of the BSE
is chosen to be frequency independent \cite{Wstatic}).
Moreover $f_{xc}$ is necessarely {\it not local}, as
the expression in Fourier space depends separately
on $\bf G$ and on $\bf G'$. In fact, its nonlocality is used as the degree of
freedom which is necessary to fulfill equality (\ref{f versus W}).

On the other hand, one can try to look
directly at
$S$ instead of focusing on $f_{xc}$.
To do that we
go back to the initial Eq.(\ref{SEQ})
which we can write in a symmetric way as
\begin{eqnarray}
S = S^{(0)}(S^{(0)} - S^{(0)} K S^{(0)})^{-1}S^{(0)}.
\end{eqnarray}
Since the independent particle polarisability is
\begin{eqnarray}
  \chi^{(0)}({\bf r},{\bf r'},\omega ) =
  \sum_{n_{1} n_{2}}
  (f_{n_{1}}-f_{n_{2}})
  \frac{\Phi(n_{1},n_{2};{\bf r}) \Phi^{*}(n_{1},n_{2};{\bf r'})}
  {\epsilon^{\rm DFT}_{n_{1}} - \epsilon^{\rm DFT}_{n_{2}} - \omega },
  \nonumber
\end{eqnarray}
the equation for the two-point $S=\chi$ becomes
\begin{eqnarray}
&&
  \chi({\bf r},{\bf r'},\omega) =
  \int d{\bf r''}d{\bf r'''} \,
  \chi^{(0)}({\bf r},{\bf r''},\omega ) \cdot
  \bigg[
  \chi^{(0)}({\bf r}_{1},{\bf r}_{2},\omega )
\nonumber \\ &&
  - \int d{\bf r}_{3}d{\bf r}_{4} \, \chi^{(0)}({\bf r}_{1},{\bf r}_{3},\omega )
     v({\bf r}_{3},{\bf r}_{4}) \chi^{(0)}({\bf r}_{4},{\bf r}_{2},\omega )
\nonumber \\ &&
     +
     \sum_{n_{1} n_{2} n_{3} n_{4}}
      \frac{\Phi^{*}(n_{1},n_{2};{\bf r}_{1}) }
      {\epsilon^{\rm DFT} _{n_{2}} - \epsilon^{\rm DFT} _{n_{1}} - \omega }
\nonumber \\ &&
      {\cal F}_{(n_{1}n_{2})(n_{3}n_{4})}
     \frac{ \Phi(n_{3},n_{4};{\bf r}_{2})(f_{n_{4}}-f_{n_{3}})}
     {\epsilon^{\rm DFT} _{n_{4}} - \epsilon^{\rm DFT} _{n_{3}} - \omega
     }
  \bigg]^{-1}_{{\bf r''},{\bf r'''}}
\nonumber \\ &&
     \cdot \chi^{(0)}({\bf r'''},{\bf r'},\omega ), \label{CHI}
\end{eqnarray}
where we have used
$\Phi(n_{1},n_{2},{\bf r}) = \Phi^{*}(n_{2},n_{1},{\bf r})$,
and substituted the term $\Phi f_{xc} \Phi^{*}$ using
Eq. (\ref{f versus W}).
In other words, 
{\it for those cases where the equality (\ref{f versus W}) can be
fulfilled, i.e. in particular when a static kernel can
be found}, we have succeeded in writing the two-point TDDFT
equation in a way that exactly yields the BSE spectrum \cite{wfKSQP}.

The explicit knowledge of the kernel is actually not needed. There is
still a four-point quantity appearing, namely $\cal F$, however only in a
matrix product instead of inversions and diagonalisations, which allows
to change space in a convenient way as it is also done in recursive
inversions of BSE \cite{BBS}. Despite this
fact, in view of the ongoing discussions about the
xc-kernel it is interesting to examine some of its
features, and in particular its long-range behaviour. In fact,
for valence (v,k) and conduction (c,k+q) states,
$\Phi(v,{\bf k},c,{\bf k}+{\bf q};{\bf G}=0)$ 
goes to zero as $q$ for small $q$.
Since ${\cal F}_{(v,c),(v,c)}$ in this limit behaves as a constant,
an $f_{xc}({\bf q},{\bf G}={\bf G'}=0)$ obtained from Eq. (\ref{fxc})
must behave as $1/{q^{2}}$. There is in fact a positive
long-range contribution stemming from the QP shift of eigenvalues (as
also predicted in ref. \cite{GGG}), and a negative one
resulting from the electron-hole interaction, which is the main point
of interest here.

Using the above results, one can also understand why the TDLDA
approximation yields much worse results for the absorption than for
the loss spectra of solids.
In fact energy-loss spectra are directly related to the inverse of
the dielectric matrix $\varepsilon^{-1} = 1+v\chi$.
For the calculation of $\chi$, the kernel $f_{xc}$
is then added to $v$, which already contains a
long-range contribution $v({\bf G}=0)$.
So the presence or absence of the long range term in $f_{xc}$
does not necessarily show up. However,
in the case of absorption spectra, one can show
\cite{HankeSham} that the macroscopic dielectric function is given by
\begin{eqnarray}
   \varepsilon_{\rm M}(\omega) =
   1 - \lim_{{\bf q}\rightarrow 0}
   [
   v(q) \bar\chi_{{\bf G}={\bf G'}=0}({\bf q},\omega)
   ]
   ,
\end{eqnarray}
where $\bar\chi$ has been calculated in a Dyson-like equation
such as (\ref{SEQ})
using the same
$f_{xc}$, but added to a coulombian $\bar v$ which does not contain
the long range term, i.e. $v({\bf G}=0)$ is set to zero. Obviously in that
case,
a neglect of the divergence in $f_{xc}$ makes
an essential difference.


We can carry this
discussion about the long-range electron-hole interaction term
further by a) assuming that we absorb the first, positive
contribution in the energy shift of our starting $\chi^{(0)}$ (since
anyway we do not know the eigenvalues of the exact
xc-potential which would go along with the exact
kernel) and
b) supposing that we have a system where the
long-range term is completely dominating the rest of the
xc
contribution, namely, where we can approximately write
$f_{xc}({\bf q},{\bf G},{\bf G'}) = 
-\delta_{{\bf G},{\bf G'}}\alpha/{|{\bf q}+{\bf G}|^{2}}$.
In other words, the long-range electron-hole attraction
part of $f_{xc}$ reduces the Hartree part, which is reasonable
since it stems from the exchange potential.
This approximation works best for systems with weakly bound excitons.
For a demonstration we have therefore
performed a TDDFT calculation for bulk silicon, in
the following way: first, we have determined the DFT-LDA electronic
structure. Second, we have constructed $\chi^{(0)}$, but with the
eigenvalues
shifted to the GW ones, in order to simulate the first part of the
kernel as explained above. Third, we have used 
$f_{xc}({\bf r},{\bf r'}) = -\alpha/{4 \pi |{\bf r}-{\bf r'}|}$, 
with the empirical value $\alpha = 0.2$.
The result of the TDDFT calculation for $\varepsilon_{\rm M}(\omega)$
is shown in Fig. 1. The dots are
the experimental results for the absorption spectrum 
(${\rm Im} (\varepsilon)$) measured by
Lautenschlager et al.\cite{Cardona} and the refraction index
(${\rm Re} (\varepsilon)$) measured by Aspnes and Studna \cite{Aspnes}.
The dot-dashed curve is the
result of a standard TDLDA calculation (i.e. using DFT-LDA
eigenvalues and the static short-range LDA xc
kernel). We find the well-known discrepancies with experiment.
The dashed curve is the BSE result.
Finally, the continuous curve is
the result of our approximate TDDFT calculation: it fits
almost perfectly all experimental features
in both real and imaginary parts of $\varepsilon$. 
It turns hence out that this {\it static
long-range} contribution to the kernel is sufficient to reproduce the
strong excitonic effect in a material with weakly bound excitons like
silicon.

In conclusion, we have derived a TDDFT equation from the
Bethe-Salpeter equation which should be particularly
suitable for practical applications to the absorption spectra of solids.
We have demonstrated that the static exchange-correlation kernel has a
long-range contribution stemming from
the electron-hole interaction. We have explained why
this long-range contribution is particularly important for the
absorption spectra of solids. At the example of bulk silicon, we have
shown how a very simple approximation for the kernel can yield
excellent agreement between the calculated TDDFT absorption
spectrum and experiment.

We are grateful to P. Ballone and R. Del Sole for useful discussions
and comments.
This work has been supported in part by the European Community
contract HPRN-CT-2000-00167.
Computer time has been granted by IDRIS (project 000544)
on the NEC SX5.

\onecolumn
\newpage

\begin{figure}
\caption{Silicon, optical absorption (bottom) and
refraction index (top panel) spectra.
Dots: experiment. Dot-dashed curve: TDLDA result. Dashed
curve: result obtained through the Bethe-Salpeter method. 
Continuous curve: TDDFT result using the
long-range kernel derived in this work.}
\label{FIGURE}
\end{figure}


\begin {references}

\bibitem{HKS} P. Hohenberg and W. Kohn, Phys. Rev. {\bf 136}, B864 (1964);
W. Kohn and L. J. Sham, Phys. Rev. {\bf 140}, A1133 (1965).

\bibitem{H} L. Hedin, Phys. Rev. {\bf 139}, 796 (1965).

\bibitem{ORGDA}
G. Onida, L. Reining, R. W. Godby, R. Del Sole and W. Andreoni,
Phys. Rev. Lett. {\bf 75}, 818 (1995).

\bibitem{ARDO}
S. Albrecht, L. Reining, R. Del Sole and G. Onida,
Phys. Rev. Lett. {\bf 80}, 4510 (1998).

\bibitem{BBS} L. X. Benedict, E. L. Shirley and R. B. Bohn,
Phys. Rev. B {\bf 57}, R9385 (1998).

\bibitem{RL} M. Rohlfing and S. G. Louie, Phys. Rev. Lett. {\bf 81},
2312 (1998).

\bibitem{RGK}
E. Runge and E. K. U. Gross, Phys. Rev. Lett. {\bf 52}, 997 (1984).

\bibitem{GDP}
E. K. U. Gross, F. J. Dobson, and M. Petersilka, {\it
Density Functional Theory} (Springer, New York, 1996) 

\bibitem{tdlda}
V. I. Gavrilenko and F. Bechstedt, Phys. Rev. B {\bf 55}, 4343 (1997).

\bibitem{TokatlyPankratov}
I. V. Tokatly and O. Pankratov, Phys. Rev. Lett. {\bf 86}, 2078 (2001).

\bibitem{footnote}
In the Hartree Fock limit, the diagonal element of
$\cal F$ reduces to Eq. (7) of X. Gonze and M. Scheffler, 
Phys. Rev. Lett. {\bf 82},
4416 (1999) at the resonance energy.

\bibitem{Wstatic}
The standard and succesfull way to introduce the electron-hole interaction in
the Bethe-Salpeter scheme has been done so far by the use
of a static screened interaction $W$\cite{ORGDA,ARDO,BBS,RL}
in the kernel of the BSE $F^{\rm BSE}$.

\bibitem{staticfxc}
We remark that the spirit of our scheme is to find a static $f_{xc}$
that mimics the role of the exact frequency dependent
$f_{xc}$ in the macroscopic
dielectric function, even if it is far from the true one.

\bibitem{wfKSQP}
Note that in order to obtain this simple formula, we had to assume
that KS orbitals and QP wavefunctions are equal. Otherwise, we could
write for example the BSE in the KS basis, and we would get a
complicated $\omega$-dependence instead of the simple
$(\epsilon_{n_{1}}-\epsilon_{n_{2}}-\omega )$ diagonal part
in Eq. (\ref{Chi Hexc}).

\bibitem{GGG} Ph. Ghosez, X. Gonze and R. W. Godby,
Phys. Rev. B {\bf 56}, 12811 (1997).

\bibitem{HankeSham} W. Hanke and L. J. Sham, Phys. Rev. Lett.
{\bf 43}, 387 (1979).

\bibitem{Cardona}
P. Lautenschlager et al., Phys. Rev. B {\bf 36}, 4821 (1987).

\bibitem{Aspnes}
D. E. Aspnes and A. A. Studna, Phys. Rev. B {\bf 27}, 985 (1983).

\end{references}
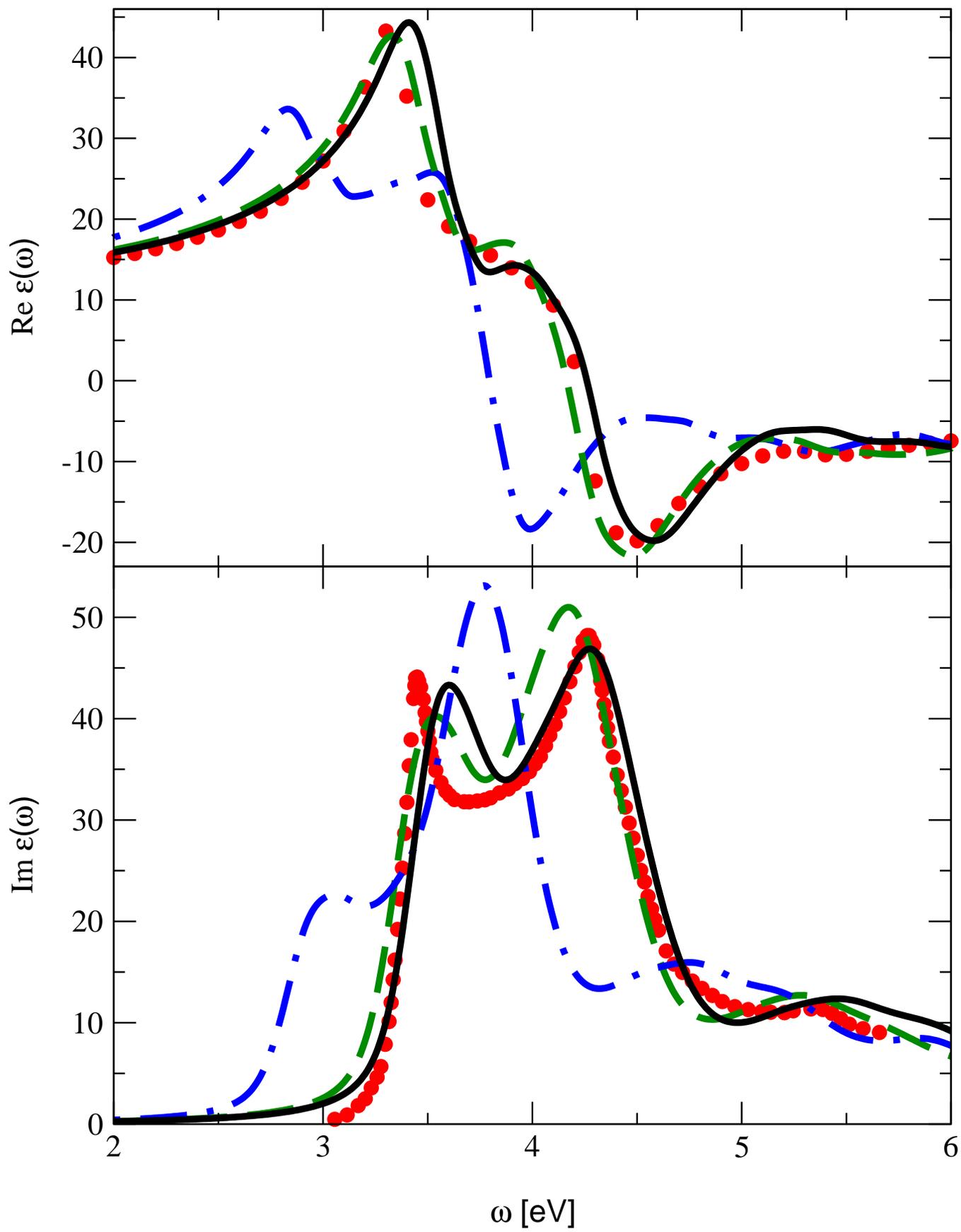
\end{document}